\input{aipcheck}
\documentclass[
    ,final            
  ,numberedheadings 
]
  {aipproc}

\layoutstyle{6x9}
\begin{document}
\title[Chiral tensors]{Chiral Lagrangians with tensor sources\thanks{Talk given at the 4th International Worshop on Quantum ChromoDynamics, Theory and experiment, June 16-20, 2007. Martina Franca - Valle d'Itria - Italy. IFIC-07-54, FTUV-07-0920. To appear in the proceedings.}}



\author{V.~Mateu}{
  address={Departament de F\'isica Te\`orica, IFIC, Universitat de Val\`encia-CSIC\\ Apt. Correus 22085, E-46071 Val\`encia, Spain. },
}

\begin{abstract}
The implementation of tensor sources in Chiral Lagrangians allows the computation of Green functions and form factors involving tensor currents, that is, quark bilinears of the form $\bar{q}_i\,\sigma^{\mu\nu}\,q_j$. Whereas only four new terms show up at $\mathcal{O}(p^4)$, we find around a hundred of them at $\mathcal{O}(p^6)$. So it becomes essential to ensure that this set o operators is indeed minimal and non--redundant ({\it i.e.}, it is a basis). We discuss two phenomenological applications in the context of vector meson resonances and the radiative pion decay.
\end{abstract}

\maketitle

\section{Introduction}
Chiral perturbation theory ($\chi$PT for short) is regarded nowadays as the effective field theory dual to QCD at very low energies. It describes the dynamics of the pseudo--Goldstone bosons ($\pi,\, K,\,\eta$) in an expansion in powers of external momenta and masses. It incorporates the assumed QCD symmetries such as parity, charge conjugation and, most important, the spontaneous breakdown of chiral symmetry.

The idea of building the chiral Lagrangian was pioneered by Weinberg \cite{Wei} and systematized by Gasser and Leutwyler \cite{Gass1}, who prepared the machinery for the computation of loop diagrams [$\mathcal{O}(p^4)$ in Weinberg power counting scheme]. One of the most interesting aspects of Gasser and Leutwyler formulation was the introduction of external static sources, rendering Green functions and form factors by functional derivation. They accounted for the Dirac bilinears  ${\bar{q}}\,\gamma_{\mu}\,q$, ${\bar{q}}\,\gamma_{\mu}\,\gamma_5\,q$, ${\bar{q}}\,q$ and ${\bar{q}}\,i\,\gamma_5\,q$. 

Even though the Chiral Lagrangian has been pushed to the two--loop order \cite{Fea,Bij} [$\mathcal{O}(p^6)$], no attempt has been made to include more external sources. For the case of quark bilinears (the ones interpolating meson fields) there is only one additional source to be considered, as the Dirac algebra is closed once the tensor current $q\,\sigma^{\mu\nu}\,q$ is added to the set previously considered. The lack of interest on this last source is somehow justified since, as opposed to the other Dirac currents, it is not realized in the Standard Model (SM). However its phenomenology is far from uninteresting~: QCD sum rules \cite{Craigie:1981jx} and Lattice \cite{Bec} study Green functions coupled to tensor currents in an effort to get insight on the dynamics of vector--meson resonances.

Finally, in certain models beyond the SM, tensor interactions arise naturally \cite{Mateu:2007tr}. One has then to calculate hadronic matrix elements of tensor currents. In our work \cite{Cata:2007ns} we provide the QCD based framework for such studies.

\section{QCD in the presence of external sources}

The external field method was first introduced by Leutwyler \cite{Leut}, and makes easier the computation of QCD Green functions and form factors. It enforces chiral Ward identities exactly by promoting the global chiral symmetry to a local one. The basic idea is to introduce in the QCD action a set of local fields coupled to the QCD currents~:
\begin{equation}
 \!Z[v_{\mu},a_{\mu},s,p,{\bar{t}}_{\,\mu\nu}]=\!\!\int\! {\cal{D}}{\bar{q}}\,{\cal{D}}q\,{\cal{D}}G_{\mu} \,{\mathrm{exp}}\!\left[i\!\int\! d^4x\,\left\{ {\cal{L}}_{QCD}^0+{\cal{L}}_{ext}(v_{\mu},a_{\mu},s,p,{\bar{t}}_{\mu\nu})\right\}\right],
\end{equation} 
where ${\cal{L}}^0_{QCD}$ is the massless QCD Lagrangian and
\begin{equation}\label{sources}
{\cal{L}}_{ext}\,=\,{\bar{q}}\,\,\gamma_{\mu}\,(v^{\mu}+\gamma_5\,a^{\mu})\,q-{\bar{q}}\,(s-i\,\gamma_5\,p)\,q+{\bar{q}}
\,\sigma_{\mu\nu}\,{\bar{t}}^{\,\mu\nu}\,q\, ,
\end{equation}
where $v_{\mu}$, $a_{\mu}$, $s$, $p$ and ${\bar{t}}^{\,\mu\nu}$ are hermitian matrices in flavor space. 
The vector and axial-vector external fields are chosen to be traceless in flavor space, but the rest of them will in general have a non-vanishing trace. 
For making the chiral symmetry transformations manifest, one uses the left-- and right--handed fermion fields. Then the tensor source is split as \cite{Cata:2007ns}
\begin{equation}
{\bar{q}}\,\sigma_{\mu\nu}\,{\bar{t}}^{\,\mu\nu}q\,=\,{\bar{q}}_L\sigma^{\mu\nu}t^{\dagger}_{\,\mu\nu}q_R\,+\,{\bar{q}}_R\,\sigma^{\mu\nu}t_{\,\mu\nu}q_L\, ,
\end{equation}
with
\begin{equation}
\!\!\!t^{\,\mu\nu}=P_L^{\mu\nu\lambda\rho}\,{\bar{t}}_{\,\lambda\rho}\,,\quad \!
P_R^{\mu\nu\lambda\rho}=\frac{1}{4}(g^{\mu\lambda}g^{\nu\rho}-g^{\nu\lambda}g^{\mu\rho}+i\,\varepsilon^{\mu\nu\lambda\rho})\, ,\quad\!
P_L^{\mu\nu\lambda\rho}=\left( P_R^{\mu\nu\lambda\rho}\right)^\dagger\!\!\!\!\!.\label{chiral-projector} 
\end{equation} 
It is not difficult to check that $P_{L\,R}$ are chiral projectors, and then $t_{\,\mu\nu}$ and $t^{\,\dagger}_{\mu\nu}$ are the left-- and right--handed projections of the tensor field ${\bar{t}}_{\,\mu\nu}$. So the tensor source is its own chiral partner, as one expects from the fact that $\gamma_5\,\sigma^{\mu\nu}$ is not independent from $\sigma^{\mu\nu}$.

We want ${\cal{L}}_{ext}$ to have the same symmetries as ${\cal{L}}^0_{QCD}$. Moreover we want chiral symmetry to hold even at a local level, that is, invariance under the transformation $q_{L\,R}\to g_{L\,R}(x)\,\,q_{L\,R}$, with $g_{L\,R}(x)$ $SU(n_f)_{L\,R}$ matrices. We cannot impose local invariance on ${\cal{L}}_{ext}$ and ${\cal{L}}^0_{QCD}$ separately~: it must be done on ${\cal{L}}_{ext}+{\cal{L}}^0_{QCD}$, being the $\ell_\mu=v_\mu+a_\mu$ and $r_\mu=v_\mu-a_\mu$ the analogs of gauge fields. It is customary to group the scalar and pseudoscalar sources in the combination $\chi=2\,B_0\,(s+i\,p)$, being $B_0$ related to the quark condensate. Then local chiral invariance dictates the transformation properties of the external sources to be~:
\begin{eqnarray}\label{sou}
r_{\mu}\rightarrow g_R\,r_{\mu}\,g_R^{\dagger}+\,i\,g_R\,\partial_{\mu}\,g_R^{\dagger}\, ,&& 
\ell_{\mu}\rightarrow g_L\,\ell_{\mu}\,g_L^{\dagger}\,+\,i\,g_L\,\partial_{\mu}\,g_L^{\dagger}\, ,\nonumber\\
\chi \rightarrow g_R\,\chi\, g_L^{\dagger}\, ,&&
t_{\,\mu\nu} \rightarrow g_R\,\, t_{\,\mu\nu}\, g_L^{\dagger}\, .
\end{eqnarray}
\section{Chiral building blocks}

$\chi$PT is based on the spontaneous breakdown of the global $G=SU(n_f)_L\otimes SU(n_f)_R$ flavour symmetry down to its subgroup $SU(n_f)_V$. The dynamical fields $\phi^a=(\pi,K,\eta)$ are assumed to be the (pseudo--)Goldstone bosons associated to this breakdown. The general formalism for effective Lagrangians with spontaneously broken symmetries was worked out by Callan, Coleman, Wess and Zumino \cite{CWZ}, and for our case groups the $n_f^2-1$ fields in the matrix $u$ transforming under $G$ as
\begin{equation}
u(\phi^a)\,=\,{\mathrm{exp}}\left(\frac{i}{2\,F_0}\,\,\phi^a\,\lambda^a\right)\,,\quad u(\phi^a)\rightarrow g_R\,u(\phi^a)\,h^{\dagger}\,=\,h\,u(\phi^a)\,g_L^{\dagger}\, .\label{CCWZ} 
\end{equation} 
$h$ is the so called compensating field and depends on $g_{L\,R}$ and $u$ itself. Eq.~(\ref{CCWZ}) defines a non--linear representation of the chiral group $G$.

At very low energies, $\phi^a$ are the only active fields and thence $u$ is the only building block of the theory. Then the QCD action in this regime reads
\begin{equation}
 Z[v_{\mu},a_{\mu},s,p,{\bar{t}}_{\,\mu\nu}]\,=\,\int {\cal{D}}\,u{\cal{D}}u^{\dagger} \,{\mathrm{exp}}\left[i\int d^4x\,{\cal{L}}_{\chi}(u
;v_{\mu},a_{\mu},s,p,{\bar{t}}_{\mu\nu})\right]\,,
\end{equation} 
where ${\cal{L}}_{\chi}$ necessarily is invariant under the same set of transformations as ${\cal{L}}_{ext}+{\cal{L}}^0_{QCD}$ is invariant.

The first step to find the most efficient set of building blocks is to manipulate $\ell_\mu$ and $r_\mu$ in such a way that the derivative is not explicit~:
\begin{eqnarray}
&& F_L^{\mu\nu}\,=\,\partial^{\mu}\ell^{\nu}-\partial^{\nu}\ell^{\mu}-i\,[\ell^{\mu},\ell^{\nu}]\, , \quad F_R^{\mu\nu}\,=\,\partial^{\mu}r^{\nu}-\partial^{\nu}r^{\mu}-i\,[r^{\mu},r^{\nu}]\, .\nonumber\\
&&u_\mu=i\left\{ u^{\dagger}(\partial_{\mu}-i\,r_{\mu})u-u(\partial_{\mu}-i\,\ell_{\mu})u^{\dagger}\right\}\quad u_\mu\to h\,u_\mu \,h^\dagger\,.
\end{eqnarray} 
Since it will prove to be very convenient to have a set of operators transforming as $u_\mu$ under $G$, we use the set introduced in Refs.~\cite{Bij,EGPR}
\begin{eqnarray}\label{listinv}
h_{\mu\nu}=\nabla_{\mu}u_{\nu}+\nabla_{\nu}u_{\mu}\, ,&&
f_{\pm}^{\mu\nu}=u\,F_L^{\mu\nu}\,u^{\dagger}\, \pm \, u^{\dagger}\, F_R^{\mu\nu}\, u\, ,\nonumber\\
t_{\pm}^{\mu\nu}=u^{\dagger}\, t^{\mu\nu}\, u^{\dagger}\, \pm \, u\, t^{\mu\nu\,\dagger} \, u\, ,&&
\chi_{\pm}=u^{\dagger} \, \chi \, u^{\dagger}\, \pm \, u\, \chi^{\dagger}\, u\, .
\end{eqnarray}
As a result, one can define a unique covariant derivative for them all, {\it{e.g.}},
\begin{equation}\label{conn}
\nabla_{\rho}X=\partial_{\rho}X+[\Gamma_{\rho},X]\, , \qquad \Gamma_{\rho}=\frac{1}{2}\left\{u^{\dagger}(\partial_{\rho}-i\,r_{\rho})u+u(\partial_{\rho}-i\,\ell_{\rho})u^{\dagger}\right\}\, ,
\end{equation} 
\section{The chiral Lagrangian}

The number of operators entering in the chiral Lagrangian is of course not restricted by symmetry. So in principle there are an infinite number of them, each one multiplied by an unknown coefficient (this is in agreement with the fact that $\chi$PT is not normalizable {\it stricto sensu}). So if one has to take them all into account in any calculation, loses all predictive power. At low energies (compared with the chiral scale $\Lambda_\chi\sim4\pi F\sim 1\, \mathrm{GeV}$) pieces with lower powers of momentum and quark masses dominate, and this allows to truncate the number of operators. This idea is systematized by the Weinberg power counting scheme \cite{Wei}.

So all of the building blocks must be endowed with a power counting, determined by the number of derivatives and masses~: $\ell_\mu$ and $r_\mu$ appear as covariant derivatives and this fixes its counting; masses count as two power of derivatives, and this dictates the counting of $s$ and $p$. Since the tensor source does not have a physical realization, its counting is arbitrary. We have then
\begin{equation}
 u \sim {\cal{O}}(p^0)\, ,\quad v_{\mu}\,,\,\,a_{\mu}\sim {\cal{O}}(p^1)\, ,\quad
\chi\,,\,\,t_{\,\mu\nu} \sim {\cal{O}}(p^2)\, ,
\end{equation} 
\

Using the tracelessness properties $\langle r_{\,\mu}\rangle=\langle \ell_{\,\mu}\rangle=\langle F_{L\,R}^{\,\mu\nu}\rangle=\langle f_{\pm}^{\,\mu\nu}\rangle=\langle u_{\,\mu}\rangle=0$ we find four new terms \cite{Cata:2007ns} at $\mathcal{O}(p^4)$ [tensor sources do not appear at $\mathcal{O}(p^2)$]
\begin{equation}\label{O4}
{\cal{L}}_4^{\chi PT}\,=\,\Lambda_1\,\langle\, t_+^{\,\mu\nu}\,f_{+\mu\nu}\,\rangle\,-\,i\,\Lambda_{2}\,\langle\, t_+^{\,\mu\nu}\,u_{\,\mu}u_{\,\nu}\,\rangle\,+\,\Lambda_{3}\,\langle\, t_+^{\,\mu\nu}\,t_{+\,\mu\nu}\,\rangle\,+\,\Lambda_{4}\,\langle\, t_+^{\,\mu\nu}\,\rangle^2\, ,
\end{equation}
where $\langle \cdots \rangle$ stands for the trace in $n_f$ flavour space. At this order there is no contact term because of the identities $t_+^{\mu\nu}\,t^+_{\mu\nu}=t_-^{\mu\nu}\,t^-_{\mu\nu}$ and $t_+^{\mu\nu}\,t^-_{\mu\nu}=t_-^{\mu\nu}\,t^+_{\mu\nu}$. 

At $\mathcal{O}(p^6)$ there are more than a hundred operators involving tensor sources, but not all of them are independent. The steps one must follow for the reduction of the number of operators were given in Ref.~\cite{Bij} and we briefly review them now~:

\vspace*{3mm}\textbf{(a) Partial integration and lowest order equations of motion}

\noindent Terms connected by partial integration leave the action invariant, and should be considered only once. The lowest order equation of motion must be used to reduce the basis (it is equivalent to a field redefinition \cite{Bij}). It reads
\begin{equation}\label{EOM}
\nabla_{\mu}u^{\,\mu}\,=\,\frac{1}{2\,i}\left(\frac{\langle \chi_-\rangle}{n_f}-\chi_-\right)\, .
\end{equation}

\textbf{(b) Bianchi identities}

\noindent There is an associated bianchi identity to the covariant derivative (\ref{conn}) that translates into one relation that allows to eliminate one operator
\begin{equation}
\nabla_{\mu}f_{+\nu\alpha}+\nabla_{\nu}f_{+\alpha\mu}+\nabla_{\alpha}f_{+\mu\nu}\,=\,\frac{i}{2}\left(\left[f_{-\mu\nu},u_{\alpha}\right]+\left[f_{-\nu\alpha},u_{\mu}\right]+\left[f_{-\alpha\mu},u_{\nu}\right]\right)\,.
\end{equation}

\textbf{(c) Isolation of contact terms}

\noindent In our basis there are linear combinations of operators which only depend on external sources. They are unphysical but however necessary to make the theory renormalizable. One has to isolate such terms, and at this order there appear only three.

\vspace*{3mm}\textbf{(c) Cayley--Hamilton relations}

\noindent The three first steps are generic for an arbitrary number of light flavours $n_f$, but the only phenomenologically interesting values are $n_f=2,\,3$. For such values additional matrix relations (based on the Cayley--Hamilton theorem) further reduce the number of independent operators. The full list of such relations can be found in Ref.~\cite{Cata:2007ns}.

\vspace*{3mm}So we end up with 117 operators plus three contat terms for $n_f$ flavours, 110+3 for $n_f=3$ and 75+3 for $2$. These operators are listed in Ref.~\cite{Cata:2007ns}. In Ref.~\cite{Haefeli:2007ty} a caveat is given, and it is claimed that the above procedure does not necessarily yield a minimal basis. However we believe that tensor sources are not affected by their considerations.

\section{Two comments on tensor sources}

So far we have restricted ourselves to the even--intrinsic sector. The odd--intrinsic sector involving external sources is characterized by being proportional to the Levi--Civit\'a symbol $\varepsilon_{\mu\nu\sigma\rho}$. Using the fact that this symbol appears on the definition of the tensor chiral projectors, Eq.~(\ref{chiral-projector}), in Ref.~\cite{Cata:2007ns} it was demonstrated that when acting on a tensor source it can be written in a form which is manifestly even--intrinsic. Then the odd--intrinsic--parity sector with tensor sources necessarily starts at $\mathcal{O}(p^8)$.

The second comment concerns the anomalous dimension of the tensor source. Since the tensor current is not conserved it must  be defined at a certain scale in QCD (much as happens with scalar and pseudoscalar currents). Since the QCD action is by construction scale invariant, LEC's accompanying tensor sources must compensate its scale dependence to make the product invariant. As a result, one should keep in mind that, besides the chirally renormalized low-energy couplings, each operator with $n$ tensor sources bears a non-vanishing anomalous dimension, namely $n\,\gamma_T$.

\section{Tensor sources in the Resonance chiral theory and the radiative pion decay}

It is well known that if one wishes to extend the energy domain of $\chi$PT resonances must be include as active degrees of freedom. The idea of constructing chiral Lagrangians with resonances fields was pioneered in Ref.~\cite{EGPR}, but the inclusion of tensor sources was first considered in Ref.~\cite{CMinprogress}. For a consistent introduction of tensor sources one has to consider spin--one resonances of $J^{PC}=1^{+-}$ quantum numbers, first included in Ref.~\cite{Zauner}. It can be shown that such resonances only manifest themselves dynamically when coupled to tensor sources \cite{CMinprogress}. However tensor sources also couple to vector--meson resonances. The definition of he vector and tensor form factors follow
\begin{eqnarray} \label{eq:fvperp}
\left\langle 0\left| \, V_{,\mu}^a \, \right|\rho_n^b(p,\lambda)\right\rangle &=& -\,\frac{1}{\sqrt{2}} \, \delta^{ab}
\, M_{V\,n} \, f_{V\,n} \, \epsilon_\mu^\lambda,\\
\left\langle 0\left| \,    T_{\mu \nu}^a \, \right|\rho^b_n(p,\lambda)\right\rangle &=& -\,
\frac{i}{\sqrt{2}} \, \delta^{ab}\,  f^\perp_{Vn} (\mu) \,\left( \epsilon_\mu^\lambda \, p_\nu-\epsilon_\nu^\lambda \, p_\mu\right), \nonumber
\end{eqnarray}
where $n$ labels the radial excitation number. In Ref.~\cite{Mateu:2007tr} a matching to the $\left\langle\, VT\right\rangle $ Green function was performed, with the assumption of minimal hadronic ansatz and in Ref.~\cite{CMinprogress} a matching to the partonic logarithm of the $\left\langle \,VV\right\rangle $ and $\left\langle \,TT\right\rangle $ Green functions is performed. The following results are quoted
\begin{equation}
\frac{f_{V\,1}^{\perp}}{f_{V\,1}}\left(1\,\mbox{GeV}\right)\,=\,0.75\pm0.14\,,\quad  \lim_{n\to\infty}\left| \frac{f_{V\,n}^{\perp}}{f_{V\,n}}\right| \sim \frac{1}{\sqrt{2}}\,. 
\end{equation} 
An intriguing result found in Ref.~\cite{CMinprogress} is that, when including in the matching the $\left\langle\, VT\right\rangle $ Green function, the above quotient must necessarily alternate its sign when moving from one multiplet to the next one.

In Ref.~\cite{Mateu:2007tr} this formalism was used to determine the tensor form factor of the pion for the $\pi\to e\,\nu \gamma$ process, obtaining
\begin{equation}
 \left\langle \, \gamma\left|\,\bar{u}\,\sigma_{\mu\nu}\,\gamma_{5}\, d\,\right|\,\pi^{-}\right\rangle =\,-\,\frac{e}{2}\, f_{T}\,(p_{\mu}\,\epsilon_{\nu}-p_{\nu}\,\epsilon_{\mu})\,,\quad f_T \,=\, 0.24 \,\pm\, 0.04\,,
\end{equation} 
and this value was used to determine the new--physics coupling to tensor currents~:  $F_{T}\,=\,(1\pm14)\times10^{-4}$.
\section{Conclusions}

We have included tensor sources in the $\chi$PT formalism and explicitly constructed the $\mathcal{O}(p^4)$ and $\mathcal{O}(p^6)$ Lagrangians. The challenging part of this construction is to find a complete and non--redundant set of operators. For the reduction of the basis we have followed standard procedures. We have demonstrated that there is no odd--intrinsic--parity sector with tensor sources at the $\mathcal{O}(p^6)$ level.

This formalism has proved to be useful for a better understanding of vector--meson spectroscopy and also for disentangling a possible tensor interaction beyond the SM in the radiative pion decay process.
\begin{theacknowledgments}
The author is supported  by a FPU contract (MEC). This work has been supported in part by the EU
MRTN-CT-2006-035482 (FLAVIAnet), by MEC (Spain) under grant
FPA2004-00996 and by Generalitat Valenciana under grant GVACOMP2007-156.
\end{theacknowledgments}

\end{document}